\documentclass{statsoc}

\usepackage{graphicx}
\usepackage{array}
\usepackage{multirow}
\usepackage[a4paper]{geometry}
\usepackage{amssymb}
\usepackage{amsmath}
\usepackage{mathrsfs}
\usepackage{graphicx}
\usepackage{natbib}
\usepackage{setspace}
\usepackage{booktabs}
\usepackage{enumerate}
\usepackage{changepage}
\usepackage{nicefrac}
\usepackage{bm}
\usepackage{pdflscape}
\usepackage{mathrsfs}
 \usepackage{url}
\usepackage{rotating}
\renewcommand{\(}{\left(}
\renewcommand{\)}{\right)}

\newcommand{\bfx}{{\mathbf{x}}}
\newcommand{\bfz}{{\mathbf{z}}}
\newcommand{\bfbeta}{{\boldsymbol{\beta}}}

\newcommand{\bfgamma}{{\boldsymbol{\gamma}}}

\newcommand{\emme}{{\mathcal{M}}}
\newcommand{\bfu}{{\mathbf{u}}}
\newcommand{\bfSigma}{{\mathbf{\Sigma}}}
\newcommand{\VSCS}{\widehat{\mathcal{M}}_{\alpha}}
\newcommand{\LBM}{\widehat{\mathcal{B}}_{\alpha}}

\title[Ranking genes by VSCS]{Ranking the importance of genetic factors by variable selection confidence sets}

\author[C.\,Zheng, D.\,Ferrari, M.\,Zhang, P.\,Baird]{Chao Zheng$^{1}$, Davide Ferrari$^{2,3}$\thanks{{\it Address for correspondence}: Department of Economics and Management, University of Bozen-Bolzano,
Piazza Universit\`{a} 1, 
39100, Bolzano, Italy.\\
E-mail: davferrari@unibz.it}, Michael Zhang$^{3,4}$ and Paul Baird$^{3,4}$}
\address{$^1$Lancaster University, Lancaster, UK\\
$^2$University of Bozen-Bolzano, Bolzano, Italy\\
$^3$University of Melbourne,Melbourne,  Australia\\
$^4$Royal Victorian Eye and Ear Hospital, Melbourne, Australia
}

\begin{document}

\begin{abstract}
The widespread use of generalized linear models in case-control genetic studies has helped identify many disease-associated risk factors typically defined as
DNA variants, or single nucleotide polymorphisms (SNPs). Up to now, most literature has focused on selecting a unique best subset of SNPs based on some statistical perspective. When the noise is large compared to signal, however, multiple
biological paths are often found to be supported by a given
dataset. We address the
ambiguity related to SNP selection by constructing a list of models -- called
variable selection confidence set (VSCS) -- which contains the collection of all
well-supported SNP combinations at a user-specified confidence level. The VSCS
extends the familiar notion of confidence intervals in the variable selection setting and provides the practitioner with new tools aiding the variable selection
activity beyond trusting a single model. Based on the VSCS, we
consider natural graphical and numerical statistics measuring the inclusion
importance of a SNP based on its frequency in the most parsimonious VSCS
models. This work is motivated by available case-control genetic data on
age-related macular degeneration,
a widespread disease and leading cause of vision loss.
\end{abstract}


 \section{Introduction}

 \subsection{Motivating AMD data and variable selection uncertainty}

Age-related macular degeneration (AMD) is a widespread condition causing blindness among adults over age 50 with estimated worldwide prevalence of 8.7\%
\citep{jonas2014global}. AMD is often referred to as a complex disease, since it is caused by the interaction of a number of genetic, environmental and
lifestyle factors, many of which have not yet been discovered. The motivating
dataset in this paper consists of case-control categorical
measurements at 20 DNA loci called single nucleotide polymporphisms (SNPs)
published through our previous work 
\citep{amd2013seven}. SNPs are substitutions of a single nucleotide ($A$ = Adenine, $T$=
Thymine, $C$=Cytosine, $G$=Guanine) at a specific position on the genome. An
understanding of their role in
disease is sought in order to better diagnose, predict disease progression and personalize treatment regimens for patients.

There are a wealth of methods in the literature of model selection for selecting a single combination of SNPs. Examples in
genetics include approaches based on information-theoretical
criteria, Bayesian  and frequentist sparsity-inducing penalization approaches; e.g., see
\citet{frommlet2012modified, wen2015bayesian, park2008penalized, wu09, ayers2010snp, li2014fast}.
Although the usual goal of model selection is to find a single optimal model, in the presence of
pronounced noise relative to signal, multiple models may  be well supported 
by a given dataset. Dealing with this model ambiguity can be challenging in 
 diseases
such as AMD due the many 
alternative pathways (combinations of SNPs) potentially explaining disease 
and modest signal-to-noise ratio.

Focusing exclusively on a single selected model, however, implies loss of information from different perspectives. First, alternative explanations of
disease etiology are tossed away -- although these may be scientifically plausible and supported by the data. Second,  the usual 
standard errors for a given selected model (e.g. regression model) fail to describe the variable selection uncertainty, so they cannot be directly used for 
establishing if a SNP subset is superior to other subsets. For similar reasons, ranking importance of SNPs based on the size of estimated coefficients and their
standard errors (e.g. z-scores) may be misleading. One might argue that ranking can be achieved by marginal measures of association (e.g. chi-square or
exact Fisher tests), but this would ultimately fail to capture the role of SNP
combinations, which is a key goal when studying complex diseases.

\subsection{A frequentist characterization of variable selection uncertainty}

This paper proposes to resolve   issues related to model ambiguity in the above AMD data by constructing a set of plausible SNP combinations,
which we call a variable selection confidence set (VSCS). We begin by taking $p$ SNPs where $p$ is typically much smaller than the sample size, $n$. For
instance, genome-wide scans and scientific considerations currently suggest that 
20 genetic loci could play a role in AMD \citep{amd2013seven}. Alternatively, 
some conservative variable selection method is applied to reduce the number of 
SNPs. Such a screening should be conservative in the sense that    
the set of selected SNPs should contain the relevant SNPs (but it is also 
allowed to contain some irrelevant ones as well). Then, within a
GLM framework, we compare all the models nested  in the full model with $p$ predictors using a likelihood ratio test (LRT) at a given
significance level $\alpha$. The final VSCS is then constructed by retaining the models not rejected by the LRT screening procedure.

An important consequence of the above LRT screening is that the resulting VSCS is guaranteed to contain the ``true model'' based on the available data with a
probability of at least $1-\alpha$ as the sample size diverges. By analogy 
with the frequentist confidence intervals for parameter estimation, a
VSCS should be regarded as a set of plausible models at the $100(1-\alpha)\%$ 
given confidence level. Under
standard conditions required to ensure the asymptotic distribution of the LRT 
screening for GLMs, the VSCS models  include the terms in the true model 
with large probability in large samples. Hence, the 
frequency of the 
selected SNPs and their combination in the VSCS models is 
expected to reflect their importance in relation to the response in a principled 
way that goes beyond just trusting a single model selected by some rule.

\cite{ferrari15} develop a procedure to 
construct variable selection confidence sets for regression models by F-testing 
and study its properties. \cite{zheng2018model} extend this methodology in the 
context of general likelihood functions. Previously, \cite{Hansen11} propose the 
model confidence sets
building upon step-down methods for multiple hypothesis testing (e.g., see 
\cite{lehmann06}). Differently
from their approach we focus specifically on LRT for GLMs in order to obtain an
asymptotic coverage probability for the globally  optimal model, which is
difficult to obtain when starting from an arbitrary list of models as in 
\cite{Hansen11}. In  
treatment regimes (DTR), one rule is 
typically selected for each stage of patient treatment. Instead of trying to 
find a single optimal DTR, \cite{wu2016set} 
construct a set of DTRs such that the optimal DTR is contained in such a set with a desired probability. 

In this paper, we apply the methods in \cite{zheng2018model}
in the context of the logistic regression model to analyze the AMD case-control 
genotype data. The empirical findings in this paper and new simulation results 
confirm our previous theoretical results. Further, in this paper we also 
explore a new way to combine the most parsimonious models in the VSCS. Stable 
ranks for SNPs combinations in 
relation to AMD and a central model in our VSCS is obtained by combining SNP 
predictors appearing with relatively high frequency in the most parsimonious 
VSCS models -- the so-called lower boundary models (LBMs).

A popular approach for assessing model uncertainty is
Bayesian  model averaging (BMA), which places prior distributions on the 
parameters along with a prior over model space (e.g., see 
\cite{bondell2012consistent} and references therein). While 
both BMA and VSCS methodologies enable one to assess the model uncertainty and 
variable importance, currently we do not have 
a well-developed 
frequentist equivalent to BMA. The two methods carry out model uncertainty 
evaluation
using fundamentally different conceptual frameworks.   Whilst our 
approach does 
not require any prior information and relies entirely on frequentist 
arguments, BMA requires the availability of a 
suitable prior distribution on the model space, which is often hard to 
obtain. Thus, the proposed VSCS methodology  fills 
the missing gap in this literature by providing a frequentist solution to the 
model uncertainty problem and it is not meant to compete directly with the 
Bayesian approach.

The rest of the paper is organized as follows. In Section 2, we present the main
VSCS methodology for generalized linear models, discuss the notion of lower
boundary models (LBMs) and introduce natural statistics based on the VSCS
to rank predictors' inclusion importance. In the same section, we propose a
model-combining strategy to obtain a single representative model
based on the inclusion importance ranks of predictors.  In Section 3, we
study the finite sample properties of our methods using simulated
genotype data. In Section 4, we apply the new methodology and study the AMD
Gene Consortium case-control genotype data. In Section 5, we conclude and give
final remarks.

\section{Methods}

\subsection{Setup: Generalized linear models}

Generalized linear models (GLMs) play an important role in
many fields of empirical research. In genetics, due to their
flexibility in modelling the relationship between predictors and a function of a
response variable,  GLMs have become popular tools to investigate the
association between SNPs and phenotype (e.g. disease occurrence). Let $Y_i$,
$i=1,\dots, n$, be independent phenotype measurements which are
assumed to follow a distribution from an exponential family with mean $\mu_i =
E(Y_i)$. Let $\bfx_i=(x_{i1}, \dots,
x_{ip})^T$, $i=1,\dots, n$, be $p$-dimensional vectors of SNP
covariates. Specifically, $x_{ij} = AA, Aa$ or $aa$, where the letters ``$A$''
and ``$a$'' represent one of the nucleotide bases in $\{A, T, C, G\}$.  Other
covariates representing demographic attributes of patients (e.g. gender,
ethnicity, etc.) are collected in  $q$-dimensional vectors,
$\bfz_{i} = (z_{i1}, \dots, z_{iq})^T$, $i=1,\dots, n$.

Let $g(\cdot)$ be an invertible linearizing link function mapping the
expectation of the response variable to
the predictors as follows
\begin{equation} \label{model1}
g(\mu_i) =  \eta_i = \beta_0 +\bfbeta^T\bfx_{i} + \bfgamma^T \bfz_{i}, \  \ i
=1, \dots, n,
\end{equation}
where $\beta_0 \in \mathbb{R}^1$, $\bfbeta=(\beta_1, \dots, \beta_p)^T \in \mathbb{R}^p$, and $\bfgamma=(\gamma_1,
\dots, \gamma_q)^T \in \mathbb{R}^q$ are  model parameters.
The intercept $\beta_0$ and the term $\bfgamma^T \bfz_{i}$ are always included
in the linear predictor (\ref{model1}), so the
main focus is on selecting subsets of SNP predictors.
Particularly, we
assume that some of the
coefficients in $(\beta
_{1}, \dots ,\beta _{p})$ are  zero and denote by $m^{\ast } \subseteq \{1,
\dots, p \}$  the
set of indexes of all non-zero terms in the true model.  The full model
containing all $p$
predictors is denoted by $m_f$ and the set of all feasible
models denoted by $\mathcal{M}$ is defined by taking all the nested models
 in $m_f$. We
assume that the sample size $n$ is sufficiently large so that estimates
$(\hat{\beta}_0, \hat{\bfbeta}, \hat{\bfgamma})$ can be obtained  by
standard likelihood  methods. 

The methods discussed in the rest of this section are rather general and can be applied to any GLM
 as long as the conditions ensuring asymptotic normality of
$(\hat{\alpha}, \hat{\bfbeta}, \hat{\bfgamma})$ are met. However, due the
nature of the motivating AMD data application in this paper,  both numerical
experiments and real data analysis in
Sections  \ref{sec:MCsimulations} and \ref{sec:real_data} will focus on the
logistic regression model
where $Y_i$ is a binary response representing presence/absence of AMD and  $E(Y_i)
= P(Y_i = 1) = \mu_i$ represents the probability of disease for the $i$-th
subject. The relationship between response and  covariates is modelled by the
usual logit link function
$\log(\mu_i) - \log(1-\mu_i) = \eta_i$, where $\eta_i$ is the linear predictor in 
 (\ref{model1}).

\subsection{VSCS construction by likelihood ratio testing} \label{sec2.2}

Given observations $ \{ (Y_i, \bfx_i, \bfz_i), i=1,\dots, n\}$, our
main interest is to construct a  set of
models, $\VSCS$,  satisfying
$ P(m^{\ast }\in \widehat{\mathcal{M}}_{\alpha}) \geq 1-\alpha
$ as the sample size increases, where $0<\alpha <1$ is a  user-defined constant 
and $m^\ast$ is the true
model.
The set $\widehat{\mathcal{M}}_\alpha$  is called a variable
selection confidence set  and all the models in $\VSCS$ may be
regarded as plausible at the user-specified $100(1-\alpha)\%$
confidence level (e.g., 95 or 99\%) \citep{ferrari15}. To obtain
$\VSCS$, we compare  the full model $m_f$  to a candidate
model $m$ nested in $m_f$  by the
likelihood ratio test (LRT) statistic
\begin{equation} \label{LR}
D_n(m) = 2 \left\{ \ell_n(m_f) - \ell_n(m) \right\},
\end{equation}
where $\ell_n(m)$ is the log-likelihood function evaluated at the maximum
likelihood estimates for model $m$. Under the null hypothesis that the model $m$ contains the true model, $D_n$ follows   a central
chi-square distribution
with $p-p_m$ degrees of freedom for large $n$.

The candidate model $m$ survives the LRT evaluation  if $D_n(m) < \chi^2(\alpha;
p-p_m)$, where $\chi^2(\alpha; \nu)$ denotes the upper $\alpha$-quantile for a
chi-squared  distribution with $\nu$ degrees of freedom, and $p$ and $p_{m}$ are the number of SNP
predictors in  $m_{f}$ and
$m$, respectively.  Then we define the $(1-\alpha)\%$-VSCS  by all the models surviving the LRT evaluation:
\begin{equation}\label{Gammahat}
\VSCS = \left\{ m \in \emme:  D_n(m) < \chi^2(\alpha, p - p_m)
\right\}.
\end{equation}
The full model $m_f$ is included in $\widehat{\emme}$ by default.
 If a model does not survive
the
LRT evaluation, then such a model is considered overly risky, in the sense that
it might miss relevant predictors  at the confidence level $100(1-\alpha)\%$.

A direct consequence of this  procedure is that
the resulting VSCS $\VSCS$
includes the true model with large probability as $n
\rightarrow \infty$. Specifically, if $m_f \neq m^\ast$ the VSCS
satisfies 
\begin{equation} \label{exthm1}
\lim_{n \rightarrow \infty} P\left( m ^{\ast }\in \VSCS \right) \geq
1-\alpha,
\end{equation}%
and in the special case where $m_f = m^\ast$, we have $\lim_{n \rightarrow
\infty} P( m^{\ast }\in \VSCS) =1$. This property follows
directly  from the well-known convergence in distribution of  the likelihood
ratio statistic $D_n$  to the central chi-square distribution with
$p-p_m$ degrees  of freedom under the null hypothesis that the smaller candidate
models are the true models.

The motivating data analysis problem in this paper allows us to compute
the VSCS by an exhaustive search since the number of SNPs in the full model is
relatively small ($p=20$). In other applications, however, the number of SNP 
predictors, $p$,  can be 
much larger than the  number of observations, $n$.  Following 
\cite{ferrari15}, in such cases we  suggest to carry out a preliminary variable 
screening to drop
unimportant predictors and reduce their number to be less than
the sample size. Consider a variable selection 
method $\psi$ yielding a reduced full model $m_{\psi}$. The size of such a  
reduced model is typically substantially smaller than $n$. For
example, in the sure independence screening procedure of \cite{ 
fan2010sure} based on marginal
associations, the prescribed size is of order $n/\log(n)$. 
Treating now $m_\psi$ as the full model, we can find the VSCS 
$\widehat{\mathcal{M}}_{\psi}$
as described previously. Alternatively, one may consider a variable 
selection methods suitable for the $p>n$ setting (e.g., 
\cite{park2007l1, xie2009scad}) producing a model with a choice of a tuning 
parameter. For our purposes, such a tuning parameter should be 
selected conservatively so that the reduced model is likely to not miss the 
true 
predictors, whilst it may include other noise variables. If the resulting model 
$\psi$ is over-consistent, i.e. with probability 
going to 1 the set of
selected predictors contains
all the predictors in the true model, then 
$
\underset{n \rightarrow \infty}{\liminf} P\left( m^{\ast }\in 
\widehat{\mathcal{M}}_{\psi}\right) \geq 1-\alpha$. This result follows 
directly from the asymptotic coverage of $1-\alpha$ of the VSCS 
\citep{zheng2018model} . For 
this to hold, it is  required that the screening is done based on data 
from a 
previous study or using a portion of the data at hand. If the sample size is 
too 
small and there is no previous data, variable screening may be done with the 
full data, although there might be a bias due to reuse of the same data for 
both variable screening and VSCS construction.

\subsection{The lower boundary models} \label{Sec:LMBs}

Without additional assumptions on the true model, the VSCS
can be  large because many models -- roughly $2^{p-p_{m^\ast}}$ -- containing the
true model plus unimportant terms survive the LRT screening. We
address the potential largeness of the VSCSs by focusing on a smaller but very
informative subset
of the   VSCS -- the set of lower boundary models (LBMs) --
hereafter denoted by $\LBM$. The lower boundary model set, $\LBM$, is defined as a subset of VSCS, containing models that do not have any nested sub-models within the VSCS. In lay terms,
the LBMs can be regarded as a subset of maximally parsimonious models which are
at the same time well-supported by the data.

As an illustration, we randomly generate two examples with  5  
predictors $X_1, \dots, X_5$, where the true model $m^*$ contains only $X_1$, 
$X_2$ and $X_3$. Predictors are uncorrelated in the first example and 
correlated in the second one. Table \ref{table:lbm_toy}, shows the resulting 
$\alpha$-level VSCSs, which contains 4 and 9 different models. For 
uncorrelated data, the $\LBM$ contains only $m_4$, which is the smallest model 
in $\VSCS$. In the correlated example, the $\LBM$ contains models $m_7$ and 
$m_9$, since for these models we cannot find any further nested models in 
the $\VSCS$.

\begin{table}
\caption{90\%-VSCSs and -LBMs for two examples. Data are generated from 
Model 1 in Section \ref{sec:MCsimulations}, with $n=100$, $p=5$, $k=3$ and 
using uncorrelated ($\rho=0$, Left table) and correlated predictors 
($\rho=0.75$, Right table).\label{table:lbm_toy}}
\centering
\hspace{-2cm}\begin{tabular}{l|ccccc|c}
\hline
$\VSCS$&$X_1$&$X_2$ & $X_3$&$X_4$&$X_5$ & p-value \\
\hline
$m_1 (m_f)$&\checkmark&\checkmark&\checkmark&\checkmark&\checkmark&1.00\\
$m_2$      &\checkmark&\checkmark&\checkmark&\checkmark&          &0.99\\
$m_3$      &\checkmark&\checkmark&\checkmark&          &\checkmark&0.24\\
\hline
\multicolumn{1}{|l|}{$m_4 (m^\ast)$}&\checkmark&\checkmark&\checkmark&          &          &\multicolumn{1}{c|}{0.45}\\
\hline
\end{tabular}
\quad\quad
\begin{tabular}{l|ccccc|c}
\hline
$\VSCS$ & $X_1$ & $X_2$& 
$X_3$& $X_4$&$X_5$&p-value\\
\hline
$m_1 (m_f)$&\checkmark&\checkmark&\checkmark&\checkmark&\checkmark&1.00\\
$m_2$      &\checkmark&\checkmark&\checkmark&\checkmark&          &0.43\\
$m_3 (m^\ast)$      &\checkmark&\checkmark&\checkmark&          &          &0.18\\
$m_4$      &\checkmark&\checkmark&          &\checkmark&\checkmark&0.10\\
$m_5$      &\checkmark&\checkmark&          &\checkmark&          &0.21\\
$m_6$      &\checkmark&          &\checkmark&\checkmark&          &0.15\\
\hline
\multicolumn{1}{|l|}{$m_7$}     &\checkmark&          &\checkmark&          &          &\multicolumn{1}{c|}{0.19}\\
\hline
$m_8 $&\checkmark& &&\checkmark&\checkmark      &0.14\\
\hline
\multicolumn{1}{|l|}{$m_9$}&\checkmark&&&\checkmark&      &\multicolumn{1}{c|}{0.22}\\
\hline
\end{tabular}
\end{table}

\begin{figure}
\caption{Variable selection confidence set ($\VSCS$) and lower 
boundary model set ($\LBM$) for increasing sample size ($n$) and different
confidence levels
(90, 95
and 99\%) for logistic regression. Top panels: Cardinality of $\VSCS$ (on the
log-scale) and $\LBM$. Bottom panels: Average Hamming distance between all the
models in  $\VSCS$   and $\LBM$ and the
true model $m^\ast$. Each point in the curves is obtained by Monte Carlo
averages based on 100 simulations from Model 1 described in Section
\ref{sec:MCsimulations}
with $(p, \rho)=(10, 0)$. }
\label{fig:con.inc}
\centering
\begin{tabular}{cc}
\centering
\includegraphics[scale=0.45]{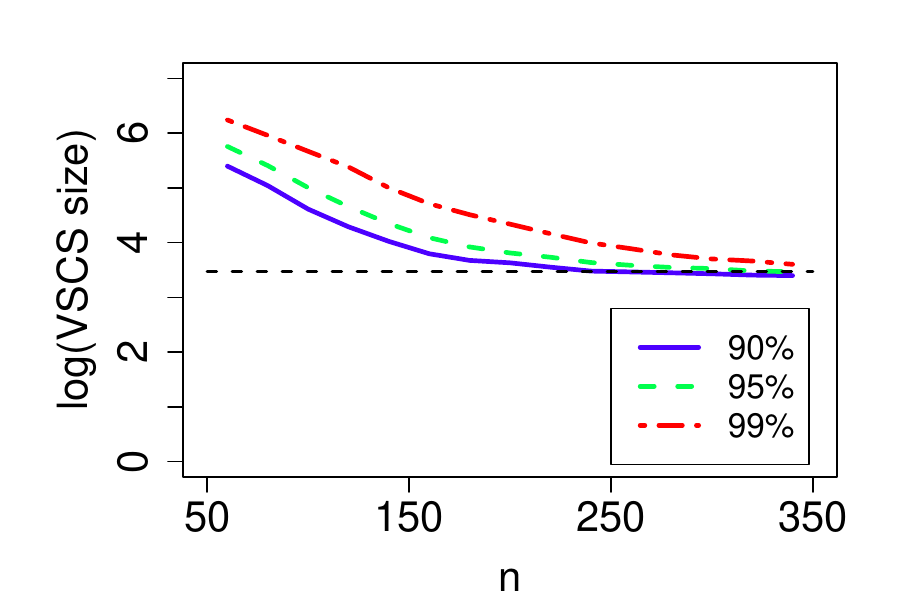}&
\includegraphics[scale=0.45]{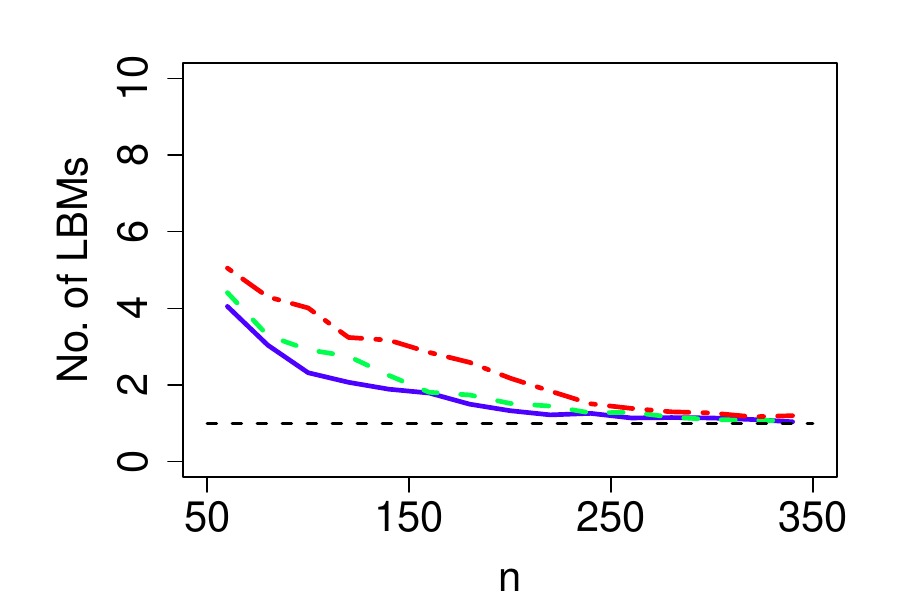}\\
\includegraphics[scale=0.45]{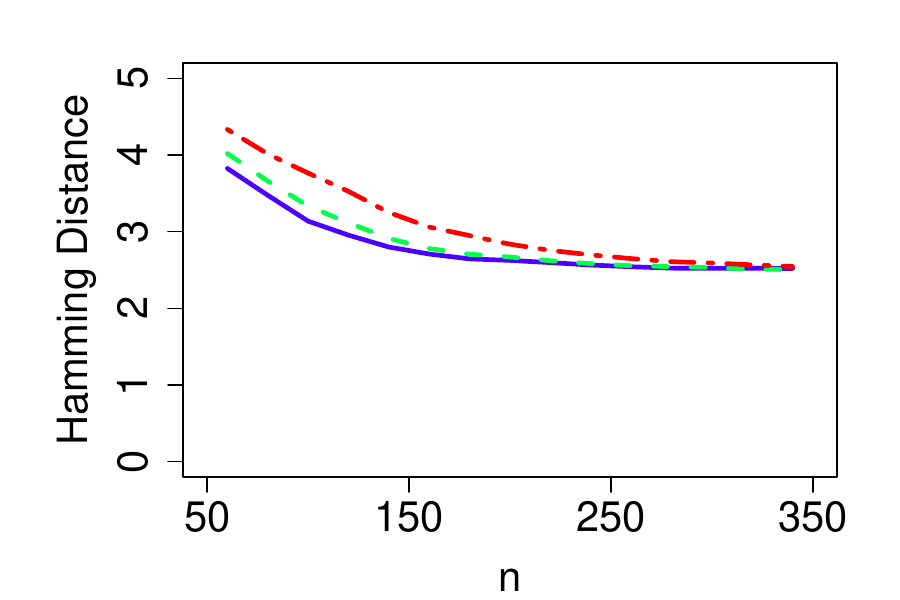}&
\includegraphics[scale=0.45]{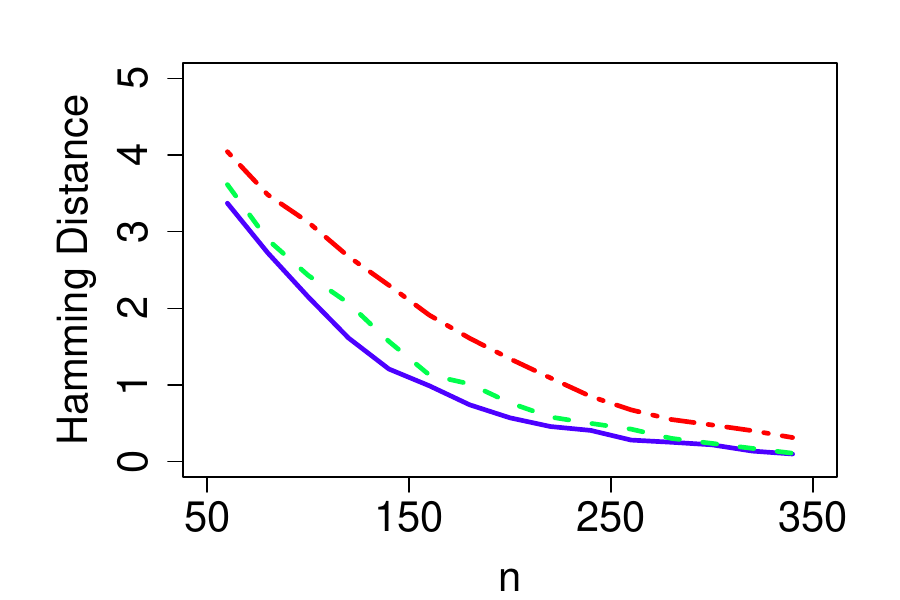}
\end{tabular}
\end{figure}

\cite{ferrari15}'s study of the LBMs in the context of linear models
shows that both the cardinality of $\LBM$ and the composition of its
constituent models carry useful information about the overall variable selection
uncertainty. It is easy to check that their results on the LBMs immediately hold
for the GLM regression
framework, when the number of predictors, $p$, is fixed and $n$ is large.
For the purposes of the current analysis, it is useful to summarize their
results by distinguishing the
following scenarios concerning the cardinality of $\LBM$:
\begin{enumerate}
\item[(i)] \textit{Zero variable selection uncertainty}. In the ideal case where we have
overwhelming information in the data,
$\LBM = \{m^\ast\}$ with large probability as $n \rightarrow \infty$  (i.e.
the LBM set contains only the true model).
This represents the ideal situation
where the sample provides us with enough information to detect all the
predictors in $m^\ast$ as they appear in the unique most parsimonious
description of the data.
\item[(ii)]\textit{Moderate variable selection uncertainty}.  A  more frequent situation in real applications
occurs when $\LBM$ contains more than 1 model, but its cardinality is much smaller
than the entire VSCS, even in the presence of pronounced noise relative to 
signal. If the
models in $\LBM$ differ by only a few predictors,
the ones appearing with relatively large frequency should be regarded as important since they
play a role in a number of parsimonious
and well-supported explanations of the data.
\item[(iii)]\textit{Strong variable selection uncertainty}. The most challenging case is when
there is too much noise compared to signal in the sample. In this situation the
size of $\LBM$ can be very large and the individual constituents of $\LBM$
may contain a   small number of predictors. This means that predictor
combinations
 are more or less picked at random by any model-selection method. In the least
favourable case, the
cardinality of $\LBM$ can be as large as ${p \choose {\lfloor {p/2} \rfloor}}$.
\end{enumerate}
The above behaviour of the LBMs is confirmed in the numerical experiments
presented in Section \ref{sec:MCsimulations}, suggesting that the
cardinality of $\LBM$ and the composition of its constituent models are helpful indicators of the total variable selection
uncertainty in a sample. As an illustration, Figure \ref{fig:con.inc}
(top panels) shows the number of models
in $\VSCS$ (on the log-scale) and $\LBM$ for different sample sizes. The graphs
are based on genotype data simulated from a logistic regression model (Model 1
in Section \ref{sec:MCsimulations} with $(p, \rho) = (10,0)$ and $m^\ast$
containing 5 predictors). Regardless of the
confidence level, when $n$ is relatively small there are
multiple LBMs quite similar to the true model, each partly overlapping with the true model. When $n$ is large, $\LBM$ tends to converge to a single
model containing only the
true predictors. Figure \ref{fig:con.inc} (bottom panels) shows
the average
Hamming distance between the models in $\VSCS$ and $\LBM$ and the true model
$m^\ast$. The Hamming distance $d_{H}(m, m^\ast)$ between individual models in
$m$ and
$m^\ast$ is
defined as the number of  different terms in $m$ compared to $m^\ast$.
As $n$ increases, the only
remaining LBM is the true model.

\subsection{Inclusion importance statistics}
\label{Sec:importance}

For linear models, \cite{ferrari15} show that in large samples true regression terms tend to appear in the LBMs with probability near 1, while
 unimportant terms appear with small probability. Our Monte
Carlo simulations in Section \ref{sec:MCsimulations} confirm this understanding
in the context of logistic regression and motivate the following
statistics to rank the importance of predictors. The inclusion importance (II)
for the $j$-th predictor is defined as
\begin{equation} \label{II}
\widehat{II}_{\alpha}(j)=\dfrac{1}{|\LBM|}{ \sum_{m \in \LBM }
I(x_j \in m ) },
\end{equation}
where  $\{x_j \in m \}$ denotes the event that the $j$-th predictor is included in
model $m$, $I(\cdot)$ is the indicator function and $|A|$
denotes the cardinality of the set $A$. If the $j$-th predictor appears in all the
boundary models then its inclusion importance is $\widehat{II}_\alpha(j) = 1$. If a predictor
appears only in a few LBMs, its II value is near zero.

It is useful to generalize the idea of inclusion importance by looking at the
joint and conditional frequencies for two or more predictors in $\LBM$. The joint
importance, or co-importance, of predictors $j$ and $k$ is defined by
\begin{align}\label{CII}
\widehat{II}_{\alpha}(j,k) = \dfrac{1}{|\LBM|} \sum_{m \in \LBM} I\(\{x_j \in m
\} \cap
\{x_k \in m \}\).
\end{align}
We remark that this measure provides information on the joint utility
of predictors in relation to the response in a way that goes beyond
simply measuring
association between predictors. For example, Figure \ref{Fig:coinclusion}
compares co-inclusion importance with pairwise sample mutual information (MI)
for the SNPs in the AMD Genotype data detailed in Section \ref{sec:real_data}.
The sample MI measures the overlap between
the distribution of two SNPs and is therefore a generic measure of association. Specifically, $MI
=  E_{\hat{p}_{jk}} \log\{\hat{p}_{jk}/(\hat{p}_{j}\hat{p}_{k})\}$, where $\hat{p}_{jk}$ and $\hat{p}_{j}$, $\hat{p}_{k}$ denote estimated joint and
marginal distributions for SNPs $j$ and $k$.
Note that while all but two SNP pairs are largely
independent,  several SNPs
co-appear with others in the LBMs with respectable frequency. This means that such pairs
contribute to a number of well-supported explanations
of the response.

If $\widehat{II}_{\alpha}(k)>0$,  the conditional importance of
predictors $j$ given $k$ is defined by
\begin{align}\label{CondI}
\widehat{II}_{\alpha}(j|k) =
\dfrac{\widehat{II}_{\alpha}(j,k)}{\widehat{II}_{\alpha}(k)}.
\end{align}
If $\widehat{II}_{\alpha}(k)=0$, predictor $k$ does not appear in the
LBMs and the conditional inclusion importance remains undefined. Given  that predictor
$k$ is in the LBM, the conditional inclusion importance statistic
(\ref{CondI}) can be used to summarize the conditional dependencies between $k$
and predictors in the LBM in terms of their role in explaining the
response variable.

The  marginal and conditional importance statistics can be displayed in a
graph as shown in Figure \ref{fig:graph} to represent what we call an
inclusion-importance network. The nodes in the graph represent SNPs
with size proportional to the marginal inclusion statistic,
$ \widehat{II}_{\alpha}(j) $. Any two nodes, say $j$ and $k$, are joined by two
directed edges with thickness proportional to
$\widehat{II}_{\alpha}(j|k)$ and $\widehat{II}_{\alpha}(k|j)$ with
arrows pointing in the direction $k \rightarrow j$ and $j \rightarrow k$,
respectively. Disconnected (or weakly connected) nodes represent conditionally
independent (or weakly dependent)
predictors, meaning that the inclusion of one  predictor is not
related to the importance of the others. Clearly, the overall degree of
connectedness and
centrality of a predictor in the graph summarize the importance of such a
predictor in relation to others for explaining the response variable.

\subsection{Combining lower boundary models} \label{Sec:ModelAggregation}

The set of lower boundary models is a summary statistic of
the variable selection uncertainty in the sample, where each LBM offers a
partial but very plausible view of the underlying data-generating process.
It is now well understood that  estimation or prediction can be improved by combining models (e.g. see
\cite{claeskens2008model} for a book-length exposition). Thus, in the same spirit of model combining methods,
we propose to aggregate the models in the set lower boundary  models to achieve a better
predictive performance than  could be obtained from any individual
constituent.

Suppose that $\LBM$ contains $k\leq p$ distinct
predictors and let $(j_1,\dots, j_k)$ denote an
arrangement of indexes in $\{1, \dots, k\}$. The LBM-aggregated model,
$\widehat{m}_{ag}$, is defined  by the index set
\begin{equation}\label{ranking}
\widehat{m}_{ag} = \left\{ (j_1,\dots, j_{\widetilde{k}}) \in \VSCS : \widehat{II}_{\alpha}(j_1) \geq \widehat{II}_{\alpha}(j_2) \geq \dots \geq
\widehat{II}_{\alpha}(j_{\widetilde{k}}),  \  \widetilde{k} \leq k\right\},
\end{equation}
where $\widehat{II}_{\alpha}(\cdot)$ is the statistic defined in (\ref{II}). Then, the
combined model
$\widehat{m}_{ag}$ contains the most useful $\widetilde{k}$ predictors as
measured by their marginal II values. Where appropriate, a
value of $\widetilde{k}$ strictly smaller than $k$ may be used to control
the complexity of the final model and avoid predictors with nonzero but very
low II value.

Although we do not offer a universally optimal rule for selecting
$\widetilde{k}$, our numerical simulations suggest that the following
step-wise strategy works well in practice. Starting
from $\widetilde{k} =1$ we progressively add candidate terms
based on their
inclusion importance rank, $\widehat{II}_\alpha(j)$, and assess the resulting
model using some model-selection criterion (e.g. AIC, BIC, etc). Then we stop if
the
given
model-selection criterion cannot
be improved and the resulting model is in $\VSCS$. In our Monte  Carlo
simulations
and real data analysis (Sections \ref{sec:MCsimulations} and
\ref{sec:real_data}),
we consider AIC and BIC as model selection criteria, and refer to the
resulting models  as AIC-$\LBM$ and BIC-$\LBM$ models, respectively.
Other selection criteria
(e.g. cross-validation) may be  used instead, but they are not explored in the
current paper.
Our numerical findings suggest that this strategy yields
parsimonious yet very informative models typically outperforming popular
model-selection methods that ignore model-selection uncertainty.

\section{Monte Carlo simulations} \label{sec:MCsimulations}

The main aims of our simulation experiments are: 1) To investigate the finite sample
coverage probability of the VSCS; 2) To study
the cardinality of the VSCS and lower boundary model set in relation to  model
selection variability; and 3) To study the performance of
the LBM-aggregation strategy described in Section \ref{Sec:ModelAggregation}.

Similarly to \cite{pan2009asymptotic} and \cite{han2012composite}, we simulate
genotype data by a  latent variable approach. We
draw independent $p$-variate vectors, $\bfu_i=(u_{i1}, \dots, u_{ip})^T$,
$i=1,\dots, n$, from a
multivariate normal distribution with zero mean vector, unit variance vector and
covariance matrix denoted by $\bfSigma$. Then we
create corresponding  genotype vectors, $\bfx_i= (x_{i1},\dots, x_{ip})^T$,
with $j$th element defined as
$$
x_{ij}=
\left\{
\begin{matrix}
Aa, &    u_{ij}< c_1, \\
AA, &    c_1 \leq u_{ij}< c_2,\\
aa, &     u_{ij} \geq   c_2,
\end{matrix}
\right.
$$
for some constants $c_1$ and $c_2$, with $Aa$, $AA$, $aa$ denoting genotype
labels. In the following simulations, we set $c_1=\Phi^{-1}(1/3)$ and
$c_2=\Phi^{-1}(2/3)$
where $\Phi(\cdot)$ is the standard normal
cdf. For each genotype vector, $\bfx_{i}$, we then generate binary responses with probability
$\pi_i = e^{\eta_i}/(1+e^{\eta_i})$, where $\eta_i$ is the linear
predictor defined in
(\ref{model1}) with $\bfgamma=(0,\dots,0)^T$ and $\beta_0=0$. The following
settings
for $\bfbeta$ and $\bfSigma$ are considered:

\vspace{0.2cm}

\noindent {\bf Model 1}: The first $k$ coefficients have the same size and the rest
are equal to 0: $\beta_j = (-1)^j$, $j = 1, \dots, k$, and $\beta_j = 0$, $j
=k+1,\dots,p$.  The elements of $\bfSigma$ follow the Toeplitz  structure:
$\bfSigma_{ij} = \rho^{|i-j|}$, $0 \leq \rho <
1$.

\vspace{0.2cm}

\noindent {\bf Model 2}: Same as Model 1, but $\bfSigma$ is a block
diagonal matrix; each block has size $(p/2) \times (p/2)$. Elements within 
each
block have covariance all equal to $\rho$, whilst between-block
covariance equal to $0$.
\vspace{0.2cm}

\noindent {\bf Model 3}: The first $k$ coefficients have decreasing size
and the rest
are equal to 0: $\beta_j = (-1)^j/j, j = 1, \dots, k \text{ and } \beta_j = 0, j
=
k+1,...,p$. Toeplitz covariance structure as in Model 1.

\vspace{0.2cm}

\noindent  {\bf Model 4}: Decreasing coefficients as in Model 2 and
block-covariance structure
as in Model 3.
\medskip

For simulation studies in this section, we fix $k$ to be $p/2$ for the above Models 1-4.

\subsection{MC Experiment 1: Cardinality/coverage of $\VSCS$ and $\LBM$ .}
Tables \ref{mc1}, \ref{mc2} and \ref{mc3} show Monte Carlo estimates for the coverage
probability, $P(m^\ast \in
\VSCS)$ (the probability that the VSCS includes true model), the cardinality of VSCS and LBM, $|\VSCS|$ and $|\LBM|$, and the average
number of predictors in the lower boundary models, $avg|\LBM|$, under different
values for the sample size ($n$), number
of predictors ($p$), correlation between predictors ($\rho$), and
significance level ($\alpha$).

For all considered models, the estimated coverage probability  is fairly
close to the nominal significance level, $\alpha$. As expected, as
$\alpha$ decreases,  the cardinality
of both VSCS and LBMs increases.   When the signal  relative to noise is small 
-- which occurs for
example when the  size of the coefficients, $\beta_j$,
decreases, or the sample size, $n$, is small -- we observe coverage probability slightly smaller than the nominal probability $1-\alpha$.  As the
correlation between variates increases, the variable selection uncertainty increases, leading to larger cardinality for both
VSCS and LBM. The coverage probability, however, is rather stable irrespective
of the correlation structure used to generate the covariates.

It is important to note that, differently from the VSCS, the cardinality of the LBM  is much smaller compared to the total number of feasible
models, $2^p$. At the same time, the LBMs contain similar information to the
whole VSCS in terms of  finding the nonzero regression terms (see
discussion in Section \ref{Sec:LMBs} and MC Example 2 in this section). This
confirms that the LBMs are informative summaries for
measuring the variable selection uncertainty  while their contained number
allows in principle to develop efficient algorithms for their discovery in
larger problems.

Finally,  from Tables  \ref{mc2} one can see that the average 
size of the individual lower boundary models is usually smaller than the size
of the true model.  Due to the effect of the variable selection uncertainty, each
LMB contains a different subset
of the true model $m^\ast$. Additional simulations
show that is quite unlikely that a LBM contains unimportant variables. This
confirms that the model combining strategy discussed in Section
\ref{Sec:ModelAggregation} represents a rather sound selection method in
its own right.

\begin{table}
\caption{
Monte Carlo estimates of the coverage probability $P(m^\ast \in \VSCS)$,under different
values for the sample size ($n$), number of predictors ($p$), correlation between predictors ($\rho$), and
confidence level. Results are obtained from 500 Monte Carlo runs. \label{mc1}}
\hspace{-3cm}\small{\begin{tabular}{p{1.4cm}<{\centering}|p{1.4cm}<{\centering}p{0.5cm}<{\centering}|p{0.5cm}<{\centering}p{0.5cm}<{\centering}p{0.5cm}<{\centering}p{0.5cm}<{\centering}|p{0.5cm}<{\centering}p{0.5cm}<{\centering}p{0.5cm}<{\centering}p{0.5cm}<{\centering}|p{0.5cm}<{\centering}p{0.5cm}<{\centering}p{0.5cm}<{\centering}p{0.5cm}<{\centering}|p{0.5cm}<{\centering}p{0.5cm}<{\centering}p{0.5cm}<{\centering}p{0.5cm}<{\centering}}
\toprule
&&~$n$~&\multicolumn{8}{c|}{100}&\multicolumn{8}{c}{200}\\
\cline{3-19}
Coverage&&~$p$~&\multicolumn{4}{c|}{8}&\multicolumn{4}{c|}{14}&\multicolumn{4}{c|}{
8}&\multicolumn{4}{c}{14}\\
\cline{3-19}
&Conf.($\%$)&~$\rho$~&0&0.25&0.5&0.75&0&0.25&0.5&0.75&0&0.25&0.5&0.75&0&0.25&0.5&0.75\\
\hline
       &90&&88.4&88.2&87.4&89.8&79.6&83.8&85.4&83.4&87.2&85.8&88.6&87.2&89.6&88.6&87.2&86.6\\
Model 1&95&&94.4&94.6&93.8&95.2&86.6&90.6&91.0&90.4&93.2&92.2&95.0&92.6&93.2&94.0&93.6&93.4\\
       &99&&98.6&98.8&98.6&97.8&94.6&96.8&97.6&97.2&98.4&98.6&99.2&99.0&98.4&98.4&99.2&99.0\\

\hline
       &90&&88.4&88.4&86.2&87.0&79.6&80.8&80.4&81.8&87.2&91.0&88.8&88.8&89.6&87.2&87.2&86.0\\
Model 2&95&&94.4&94.0&92.8&92.4&86.6&89.2&88.2&88.0&93.2&95.2&95.0&94.0&93.2&92.2&92.8&92.0\\
       &99&&98.6&97.6&98.6&97.8&94.6&96.6&95.8&96.8&98.4&98.4&99.2&98.6&98.4&97.8&98.8&98.2\\
\hline
       &90&&87.0&87.0&88.8&86.2&84.0&84.0&82.6&84.2&89.2&88.4&88.2&89.4&86.0&85.4&85.8&86.2\\
Model 3&95&&92.0&92.0&93.6&93.0&92.2&93.4&90.8&89.2&95.6&93.6&93.6&93.0&93.4&91.8&92.6&93.8\\
       &99&&97.8&98.0&98.2&99.4&98.4&99.2&97.4&97.4&99.0&98.8&98.2&99.0&98.8&98.8&98.6&98.8\\
\hline
       &90&&87.0&90.0&88.0&86.4&84.0&84.4&82.8&82.8&89.2&90.2&89.4&88.4&86.0&87.8&86.4&88.0\\
Model 4&95&&92.0&94.8&94.4&91.2&92.2&89.0&89.8&91.2&95.6&94.4&95.2&94.0&93.4&93.2&92.0&93.8\\
       &99&&97.8&98.8&99.0&98.2&98.4&97.0&97.4&97.8&99.0&99.4&99.0&98.6&98.8&98.6&97.0&98.4\\
\bottomrule
\end{tabular}}
\end{table}

\begin{table}
\caption{
Monte Carlo estimates of the cardinality of LBM sets, $|\LBM|$under different
values for the sample size ($n$), number of predictors ($p$), correlation between predictors ($\rho$), and
confidence level. Results are obtained from 500 Monte Carlo runs. \label{mc2}}
\centering
\hspace{-3cm}\small{\begin{tabular}{p{1.4cm}<{\centering}|p{1.4cm}<{\centering}p{0.5cm}<{\centering}|p{0.5cm}<{\centering}p{0.5cm}<{\centering}p{0.5cm}<{\centering}p{0.5cm}<{\centering}|p{0.5cm}<{\centering}p{0.5cm}<{\centering}p{0.5cm}<{\centering}p{0.5cm}<{\centering}|p{0.5cm}<{\centering}p{0.5cm}<{\centering}p{0.5cm}<{\centering}p{0.5cm}<{\centering}|p{0.5cm}<{\centering}p{0.5cm}<{\centering}p{0.5cm}<{\centering}p{0.5cm}<{\centering}}
\toprule
&&~$n$~&\multicolumn{8}{c|}{100}&\multicolumn{8}{c}{200}\\
\cline{3-19}
$\boldmath{|\LBM|}$&&~$p$~&\multicolumn{4}{c|}{8}&\multicolumn{4}{c|}{14}&\multicolumn{4}{c|}{
8}&\multicolumn{4}{c}{14}\\
\cline{3-19}
&Conf.($\%$)&~$\rho$~&0&0.25&0.5&0.75&0&0.25&0.5&0.75&0&0.25&0.5&0.75&0&0.25&0.5&0.75\\
\hline
       &90&&1.39&1.44&1.76&2.64&4.31&4.17&5.40&7.73&1.08&1.08&1.10&1.37&1.64&1.61&2.19&4.28\\
Model 1&95&&1.59&1.66&1.92&3.29&5.54&5.29&6.69&8.34&1.08&1.09&1.14&1.51&1.96&1.95&2.83&5.08\\
       &99&&2.07&2.20&2.76&4.61&8.49&7.64&8.62&9.78&1.14&1.16&1.34&2.11&3.07&3.14&4.70&7.74\\

\hline
       &90&&1.39&1.44&1.73&2.34&4.31&5.21&6.64&8.57&1.08&1.08&1.18&1.45&1.64&2.05&3.05&5.44\\
Model 2&95&&1.59&1.66&2.10&2.91&5.54&6.34&7.80&9.74&1.08&1.10&1.16&1.59&1.96&2.64&3.83&6.95\\
       &99&&2.07&2.26&2.80&4.14&8.49&9.51&9.67&10.32&1.14&1.26&1.36&2.10&3.07&3.95&5.83&9.43\\
\hline
       &90&&1.85&2.04&2.21&3.00&5.55&4.69&5.54&7.53&1.71&1.69&1.65&1.77&4.52&3.99&3.64&4.57\\
Model 3&95&&2.08&2.22&2.65&3.67&5.57&5.26&6.20&8.22&1.73&1.70&1.72&1.88&4.60&3.74&3.68&4.42\\
       &99&&2.70&3.03&3.89&5.07&6.49&6.92&7.89&10.27&1.86&1.69&1.72&2.29&3.83&3.28&3.87&5.35\\
\hline
       &90&&1.85&1.99&2.47&3.03&5.55&5.13&6.13&8.25&1.71&1.64&1.71&1.96&4.52&4.52&4.49&5.63\\
Model 4&95&&2.08&2.38&2.89&3.63&5.57&5.53&6.61&8.84&1.73&1.73&1.69&2.15&4.60&4.21&4.30&5.82\\
       &99&&2.70&3.44&3.93&4.96&6.49&7.23&7.97&9.92&1.86&1.72&1.88&2.62&3.83&3.88&4.64&6.83\\
\bottomrule
\end{tabular}}
\end{table}

\begin{table}
\caption{
Monte Carlo estimates of the average
number of predictors in the lower boundary models $avg(\LBM)$, under different
values for the sample size ($n$), number of predictors ($p$), correlation between predictors ($\rho$), and
confidence level. Results are obtained from 500 Monte Carlo runs. \label{mc3}}
\hspace{-1.5cm}\small{\begin{tabular}{p{1.4cm}<{\centering}|p{1.4cm}<{\centering}p{0.5cm}<{\centering}|p{0.5cm}<{\centering}p{0.5cm}<{\centering}p{0.5cm}<{\centering}p{0.5cm}<{\centering}|p{0.5cm}<{\centering}p{0.5cm}<{\centering}p{0.5cm}<{\centering}p{0.5cm}<{\centering}|p{0.5cm}<{\centering}p{0.5cm}<{\centering}p{0.5cm}<{\centering}p{0.5cm}<{\centering}|p{0.5cm}<{\centering}p{0.5cm}<{\centering}p{0.5cm}<{\centering}p{0.5cm}<{\centering}}
\toprule
&&~$n$~&\multicolumn{8}{c|}{100}&\multicolumn{8}{c}{200}\\
\cline{3-19}
$\boldmath{avg(\LBM)}$&&~$p$~&\multicolumn{4}{c|}{8}&\multicolumn{4}{c|}{14}&\multicolumn{4}{c|}{
8}&\multicolumn{4}{c}{14}\\
\cline{3-19}
&Conf.($\%$)&~$\rho$~&0&0.25&0.5&0.75&0&0.25&0.5&0.75&0&0.25&0.5&0.75&0&0.25&0.5&0.75\\
\hline
       &90&&3.85&3.70&3.45&2.83&6.42&6.03&5.63&4.74&4.15&4.16&4.06&3.81&6.98&6.96&6.73&6.19\\
Model 1&95&&3.61&3.42&3.06&2.52&6.11&5.73&5.19&4.18&4.06&4.05&3.90&3.54&6.82&6.74&6.50&5.75\\
       &99&&3.10&2.87&2.54&1.93&5.49&4.99&4.30&3.21&3.93&3.87&3.61&3.01&6.50&6.37&6.03&5.08\\

\hline
       &90&&3.85&3.59&3.33&2.70&6.42&6.13&5.60&4.64&4.15&4.10&4.05&3.68&6.98&6.87&6.69&6.03\\
Model 2&95&&3.61&3.28&3.01&2.38&6.11&5.73&5.13&4.14&4.06&3.99&3.86&3.38&6.82&6.64&6.36&5.67\\
       &99&&3.10&2.78&2.43&1.94&5.49&4.99&4.26&3.20&3.93&3.77&3.51&2.83&6.50&6.23&5.86&4.91\\
\hline
       &90&&1.97&1.75&1.67&1.54&2.40&2.21&2.14&1.96&2.52&2.36&2.04&1.71&2.98&2.69&2.21&2.19\\
Model 3&95&&1.72&1.52&1.48&1.34&2.15&1.95&1.84&1.72&2.22&2.07&1.77&1.53&2.66&2.33&1.92&1.88\\
       &99&&1.40&1.22&1.25&1.15&1.69&1.57&1.49&1.37&1.85&1.59&1.39&1.26&2.03&1.76&1.56&1.61\\
\hline
       &90&&1.97&1.78&1.66&1.61&2.40&2.33&2.28&2.10&2.52&2.28&2.13&1.85&2.98&2.73&2.51&2.28\\
Model 4&95&&1.72&1.59&1.46&1.43&2.15&2.01&1.94&1.84&2.22&2.02&1.86&1.63&2.66&2.36&2.18&2.02\\
       &99&&1.40&1.29&1.21&1.19&1.69&1.69&1.54&1.43&1.85&1.60&1.50&1.32&2.03&1.91&1.81&1.66\\
\bottomrule
\end{tabular}}
\end{table}

\subsection{MC Experiment 2: Model aggregation by importance ranking.} In our second Monte Carlo experiment, we focus on the LBMs and study
whether the occurrence of predictors in the LBM is informative on the true regression terms. To this end, we consider retaining the first
$\widetilde{k}$ regression terms, ranked according to the inclusion importance
statistic (\ref{II}) introduced in Section
\ref{Sec:ModelAggregation}. The number of terms appearing in the final selected
model, $\widetilde{k}$, is chosen by step-wise selection using AIC and BIC
scores. To assess the model-selection performance of our method, we compute
Monte Carlo estimates of the Hamming distance $d_{H}(\hat{m}, m^\ast)$ defined
by the number of different terms in a selected model $\hat{m}$
compared to the true model $m^\ast$.

\begin{table}
\caption{Monte Carlo estimates of the Hamming distance between the true model $m^\ast$ and models selected by LBM-aggregation based on AIC and BIC (AIC-$\LBM$
and BIC-$\LBM$), forward-selection based on AIC and BIC  (F-AIC and F-BIC), and LASSO, SCAD, MCP sparsity-inducing penalization approaches.
For the  LBM-aggregation method we consider 90, 95 and
99\% confidence levels. The tuning parameters for LASSO, SCAD and MCP methods are chosen by 5-fold
cross validation.  Results are based on $500$
simulations from Model 1.\label{table.hd}}
\centering
\hspace*{-1.5cm}\begin{tabular}{ccc|ccc|ccc|ccccc}
\toprule
&&&\multicolumn{3}{c|}{AIC-$\LBM$}&\multicolumn{3}{c|}{BIC-$\LBM$}&F-AIC
&F-BIC&{LASSO}&{SCAD}&{MCP}\\
$n$&$p$&$\rho$& ~$90\%$~ & ~$95\%$~  & ~$99\%$~  & ~$90\%$ ~ & ~$95\%$  ~& ~$99\%$ ~ & \multicolumn{5}{c}{}\\
\hline
{100}&{8}&  0 
&1.00&1.16&2.10&1.66&1.60&1.62&1.79&1.90&1.51&1.31&2.16 \\
                                       &~& 0.5&1.44&1.68&2.51&2.11&2.20&2.34&2.01&3.18&1.96&1.80&2.48 \\
    \cline{2-14}
  &{14}                  & 0 
&2.19&2.28&2.38&2.31&2.34&2.24&2.85&3.41&2.15&2.17&3.58\\
                                     & & 0.5 &2.90&3.04&3.65&3.33&3.53&3.70&3.48&5.75&4.15&4.12&4.48\\
      \hline
  {200}& {*}{8}& 
0&0.48&0.50&0.84&0.36&0.31&0.38&1.67&1.13&1.41&1.04&2.31\\
                                       &  & 0.5&0.57&0.66&1.58&1.04&0.90&0.93&1.63&2.12&1.38&1.08&2.59\\
      \cline{2-14}
                       &{*}{14}& 
0&1.18&1.24&1.89&1.27&1.40&1.46&2.21&1.34&2.55&1.73&4.06\\
                                         & &0.5&1.47&1.62&1.86&1.40&1.61&1.56&1.99&3.40&2.66&2.07&4.39\\
\bottomrule
\end{tabular}
\end{table}

Table \ref{table.hd} shows Monte Carlo estimates of the Hamming distance
for our LBM-aggregation methods (AIC-$\LBM$ and BIC-$\LBM$) and classic
forward-selection based on AIC and BIC scores (F-AIC and F-BIC). In addition, we
show results for other common
selection methods based on penalization approaches: least absolute selection and shrinkage operator (LASSO), smoothed clipped absolute deviation (SCAD), and
mini-max convex penalization (MCP) approaches (e.g. see \cite{fan2001variable,
zhang2010nearly}). To compute LASSO, SCAD and MCP we used the R package
\texttt{ncvreg} with tuning parameters chosen by
5-fold cross validation. The proposed LBM-aggregation methods  generally outperform
all the other selection procedures relying on a single model in most cases,
regardless of the model structure and sample size. As the sample size
increases, the accuracy of the aggregation strategy based on LBMs is
found to be significantly improved compared to single-model selection methods.

\subsection{MC Experiment 3: Comparison with Bayesian model averaging.}

In this experiment, we compare the VSCS method with Bayesian model averaging 
(BMA) (e.g. see  
\cite{claeskens2008model}).  Table \ref{table.bma} shows Monte Carlo 
estimates of the probability that the best model given by BMA (the model with 
the highest 
posterior distribution) is included in the $95\%$-level VSCS. The BMA models 
are computed using the R-package \texttt{BMA} \citep{raftery2005bma} assuming a 
uniform prior and setting the maximum ratio for excluding models in Occam's 
window as 20. The probability estimates are  not very large, meaning that the 
best BMA model varies substantially. 
Table \ref{table.bmacov} shows 
Monte Carlo estimates of the coverage 
probability for BMA models under different Occam's window 
ratios (OR) and uniform prior. The results are based on 500 Monte Carlo 
simulations from Models 1 and 3. Our results suggest that, compared to VSCS, 
that BMA 
suffers from lower coverage when the correlation between predictors increases 
or 
the number of predictors grows.

\begin{table}
\caption{Percent fraction of BMA best models included in VSCS . We set the 
maximum ratio
for excluding models in Occam's window equals to 20 and use a uniform prior to 
obtain BMA models. Results are based on 500 Monte Carlo simulations, with data generated from Model 1.
\label{table.bma}}
\hspace*{-1.5cm}
\small{\begin{tabular}{p{1.5cm}<{\centering}p{0.5cm}<{\centering}|p{0.5cm}<{\centering}p{0.5cm}<{\centering}p{0.5cm}<{\centering}p{0.5cm}<{\centering}|p{0.5cm}<{\centering}p{0.5cm}<{\centering}p{0.5cm}<{\centering}p{0.5cm}<{\centering}|p{0.5cm}<{\centering}p{0.5cm}<{\centering}p{0.5cm}<{\centering}p{0.5cm}<{\centering}|p{0.5cm}<{\centering}p{0.5cm}<{\centering}p{0.5cm}<{\centering}p{0.5cm}<{\centering}}
\toprule
&~$n$~&\multicolumn{8}{c|}{100}&\multicolumn{8}{c}{200}\\
\cline{2-18}
&~$p$~&\multicolumn{4}{c|}{8}&\multicolumn{4}{c|}{14}&\multicolumn{4}{c|}{
8}&\multicolumn{4}{c}{14}\\
\cline{2-18}
Conf.($\%$)&~$\rho$~&0&0.25&0.5&0.75&0&0.25&0.5&0.75&0&0.25&0.5&0.75&0&0.25&0.5&0.75\\
\hline
90&&0.48&0.47&0.45&0.48&0.23&0.19&0.24&0.40&0.82&0.75&0.69&0.49&0.67&0.66&0.47&0.31\\
95&&0.59&0.59&0.57&0.63&0.32&0.29&0.34&0.54&0.87&0.83&0.76&0.60&0.77&0.75&0.58&0.44\\
99&&0.79&0.81&0.81&0.86&0.52&0.50&0.57&0.79&0.93&0.90&0.89&0.81&0.89&0.88&0.79&0.68 \\
\bottomrule
\end{tabular}}
\end{table}

\begin{table}
\caption{Percent coverage probability for BMA models under different Occam's 
window 
ratios (OR) and uniform prior. Results are based on 500 Monte Carlo 
simulations from Model 1 and Model 3.\label{table.bmacov}}

\centering
\small{\begin{tabular}{c|cc|cc|cc|cc|cc}
\toprule
&&~$n$~&&\multicolumn{2}{c}{100}&&&\multicolumn{2}{c}{200}&\\
\cline{3-11}
&&~$p$~&\multicolumn{2}{c|}{8}&\multicolumn{2}{c|}{14}&\multicolumn{2}{c|}{8}&\multicolumn{2}{c}{14}\\
\cline{3-11}
&OR&~$\rho$~&0.25&0.75&0.25&0.75&0.25&0.75&0.25&0.75\\
\hline
{Model 1}&10&& 92.0&56.2&1.2&0.0&99.2&91.4&95.8&45.0\\
       &20&&96.2&70.6&9.8&0.4&99.4&95.6&98.4&60.6\\
       &50&& 99.4&86.4&39.8&10.6&99.8&98.0&99.4&78.0 \\
       &100&&99.6&91.6&61.8&27.0&99.8&99.4&99.6&85.0\\
\hline
{Model 3}&10&&13.0&5.8&0.0&0.0&20.6&4.6&0.0&0.0\\
       &20&&21.0&11.2&0.0&0.0&35.0&11.2&0.4&0.0\\
       &50&&40.6&21.6&0.0&0.0&54.8&22.4&0.6&0.2\\
       &100&&62.8&40.6&0.6&0.2&68.0&37.8&1.8&0.2\\
\bottomrule
\end{tabular}}
\end{table}

\section{Analysis of the AMD Genotype data} \label{sec:real_data}

The AMD Genotype data analysed in this section consist of measurements on
patients from either outpatient clinics at the Royal Victorian Eye and
Ear Hospital or through private ophthalmology practices in Melbourne, Australia. Control subjects were selected from the same community. The subjects
were Caucasian of Anglo-Celtic ethnic background. Patient collection and
clinical examination were undertaken as described previously in
\cite{baird2004varepsilon2} . The cohort consisted of 418 subjects with advanced AMD and 266 healthy control subjects with no AMD. For each patient, demographic
information was collected along with age and gender  and included in our models to control for potential confounding.
The 20 SNPs considered here include 14 well-replicated variants in the AMD risk assessment literature as well as 6 newly discovered variants
\citep{amd2013seven}. Genomic DNA was prepared from peripheral venous blood with genotyping of the 20 SNPs performed on the Mass Array platform (SEQUENOM, San
Diego, CA) at the Murdoch Children's Research Institute, Melbourne as previously described in \cite{gu2013rare}.

\begin{table}
\caption{Variable selection summary statistics for the AMD
Gene Consortium data: Cardinality of VSCS and LBM sets ($|\VSCS|$  and $|\LBM|$),
5-number summary statistics for the number of predictors in the LMBs, and
average Hamming distance (AHD) between pairs of LMB models, for different confidence levels.
\label{Table1:Sec4}}
 \centering
\begin{tabular}{ccccccccc} \toprule
{Conf.$(\%)$} & {$|\VSCS|$}  & {$|\LBM|$} & 
\multicolumn{5}{c}{No. predictors in LBM} &{AHD}\\
&&&  min &  Q1 & median  & Q3  & max & \\
\midrule
90  & 30827         & 43       & 6.0&   7.0&   7.0   &8.5  &10.0&5.2\\
95  & 57273         & 79       &5.0   &7.0   &8.0  &   8.0&  10.0&5.9\\
99  & 156459        &103        &4.0  &5.0 &6.0 &7.0 &10.0&6.1\\
\bottomrule
\end{tabular}
\end{table}

\subsection{VSCS, LBMs and ranks of SNPs.} Table \ref{Table1:Sec4} shows summary
statistics for the
VSCS and LBM set at the 90, 95 and 99\% confidence levels. As expected, the
cardinality of both $\VSCS$ and
$\LBM$ decrease as
$\alpha$ increases and the
cardinality for the variable selection confidence set, $|\VSCS|$, is much larger compared
to that of the LBM set, $|\LBM|$. The number of lower boundary models
is relatively large (43, 79 and 103) with a median number of SNPs
for individual LBMs ranging from 6 to 8, thus evidencing considerable model
uncertainty. We also report the average hamming
distance~(AHD) for all model pairs in the LBM computed as
$$
AHD =    {{|\LBM|}\choose{2}}^{-1} \sum_{m, m' \in \LBM :m \neq m' } d_{H}(m, m'),
$$
where $d_{H}(m, m')$ denotes the number of SNPs at which $m$ and $m'$ are different.
This quantity remains stable with
$\alpha$ and is not negligible compared to the total number of SNP
(20 SNPs), which indicates heterogeneous SNP combinations in the LBMs.
Overall, the findings in Table \ref{Table1:Sec4} show that the
variable-selection uncertainty is quite strong, despite the relatively large
sample size ($n=684$) and suggest that the data
may not be be well represented unequivocally by a single logistic regression
model selected by a method that ignores the selection-uncertainty.

\begin{figure}
\caption{SNPs ranking for the AMD gene consortium data based on the marginal
inclusion importance statistic, $\widehat{II}$, computed from $95\%$ LBM. (a)The $\widehat{II}$ for all the SNPs
(b) $\widehat{II}$ for 5 SNPs and 5 random generated predictors.}
    \label{fig:inclusion}
\centering
\begin{tabular}{cc}
\includegraphics[scale=0.50]{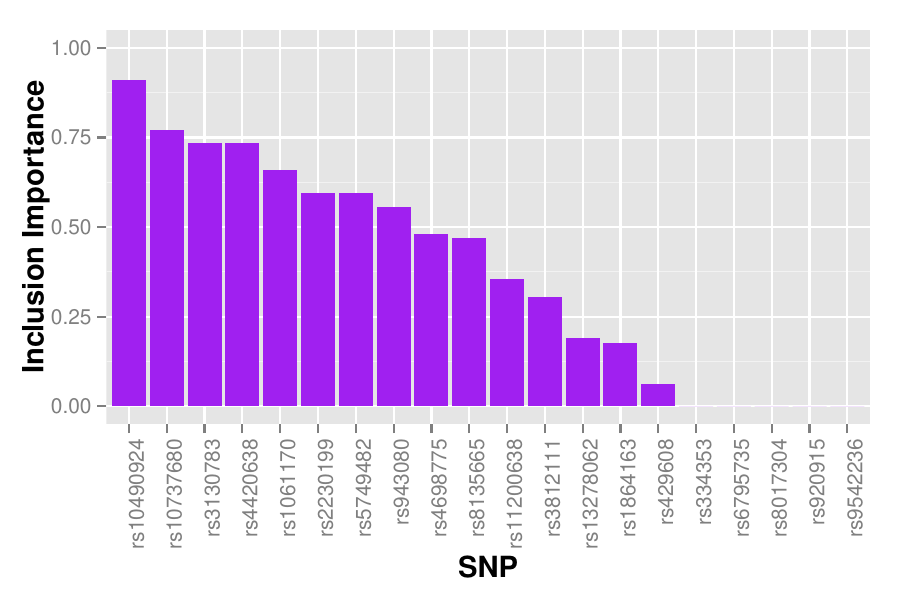}&
\includegraphics[scale=0.50]{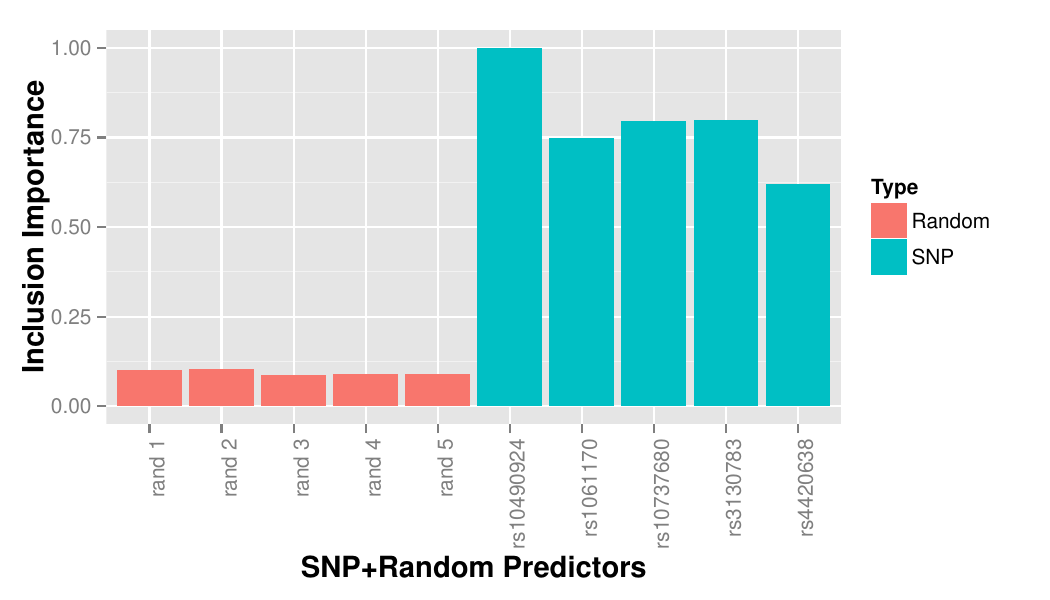}\\
(a)&(b)
\end{tabular}
\end{figure}

\begin{figure}
\caption{Conditional inclusion importance graph at the 95\% confidence level for SNPs in the AMD genotype data.  The size of the nodes and thickness of edges is
proportional to
marginal and conditional inclusion importance defined in (\ref{II}) and (\ref{CondI}), respectively.  Edges with conditional inclusion importance less than 0.7
are omitted for clarity.}
\label{fig:graph}
\begin{tabular}{c}
\centering
\includegraphics[scale=1.2]{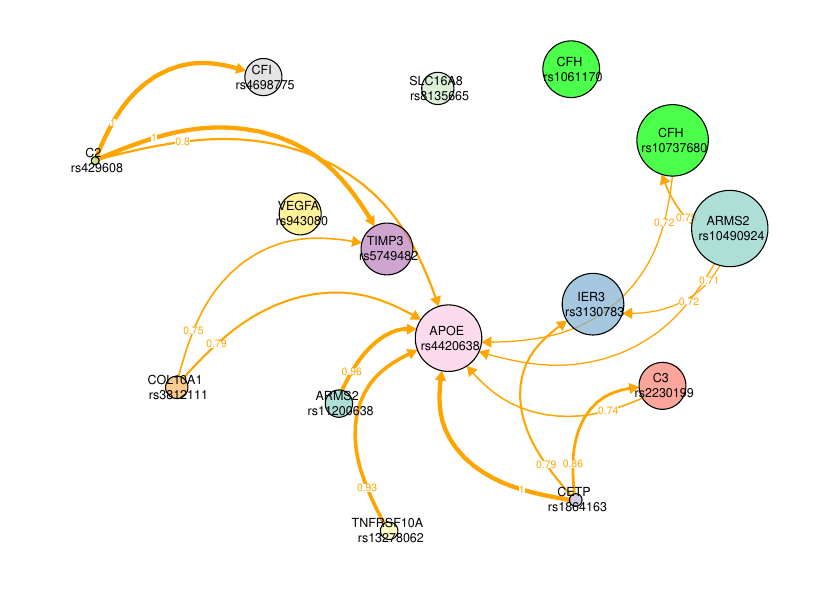}
\end{tabular}
\end{figure}

Figure \ref{fig:inclusion} summarizes SNP ranking according to the marginal
inclusion importance statistic defined in
(\ref{II}). Note that  15 SNPs exhibit non-zero importance,  about half of which
receive II values greater than 1/2. Figure \ref{fig:graph} illustrates the
conditional inclusion importance graph for SNPs in the AMD data. A
few SNPs with large marginal inclusion importance are central
in the graph (e.g. see \texttt{rs4420638},
\texttt{rs3130783}). This suggests that such SNPs may play an important role in
relation to AMD etiology in the presence of a number of other SNPs
connected in the graph. On the other hand,
certain SNPs, such as \texttt{rs10490924}, seem to be important in their
own right regardless of the presence of other SNPs.
Such SNPs lay on the
periphery of the  inclusion importance graph, meaning that they do not explain
much of the disease occurrence conditionally to the presence of other SNPs.

\begin{figure}
\caption{Co-inclusion importance at the $95\%$ confidence level and sample mutual information for pairs of
SNPs in the AMD genotype data. Left: Standardized co-inclusion
importance
$\widehat{II}_{\alpha}(j,k)/\widehat{II}_{\alpha}(j \text{ or } k)$, where
$\widehat{II}_{\alpha}(j \text{ or } k) = \widehat{II}_{\alpha}(j)+
\widehat{II}_{\alpha}(k)- \widehat{II}_{\alpha}(j,k)$ is the relative frequency
of either SNP $j$ or $k$ in the LBMs. Right: Pair-wise sample mutual information for SNPs $x_j$ and $x_k$.}
\label{Fig:coinclusion}
\centering
\begin{tabular}{cc}
Co-inclusion importance & Mutual information \\
\includegraphics[scale=0.5]{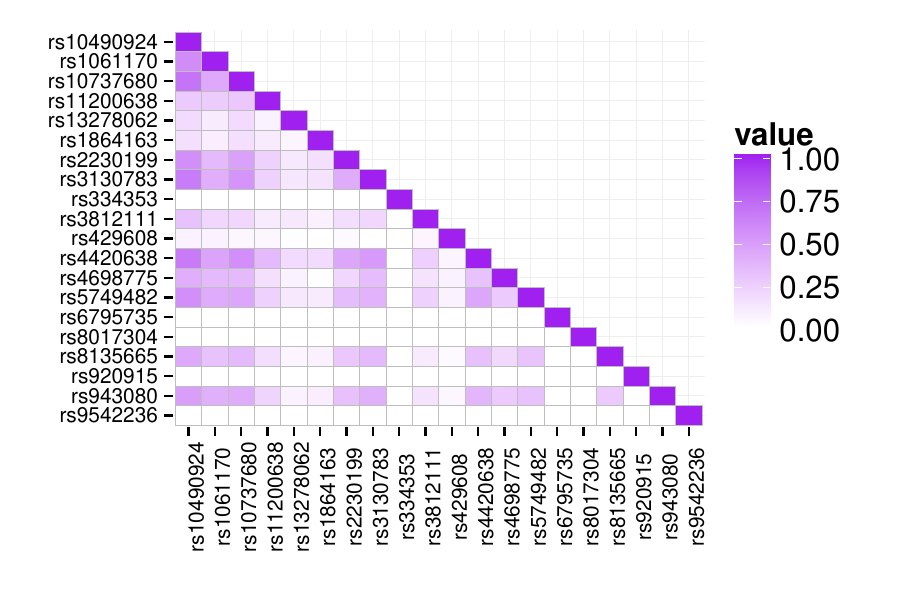}
&
\includegraphics[scale=0.5]{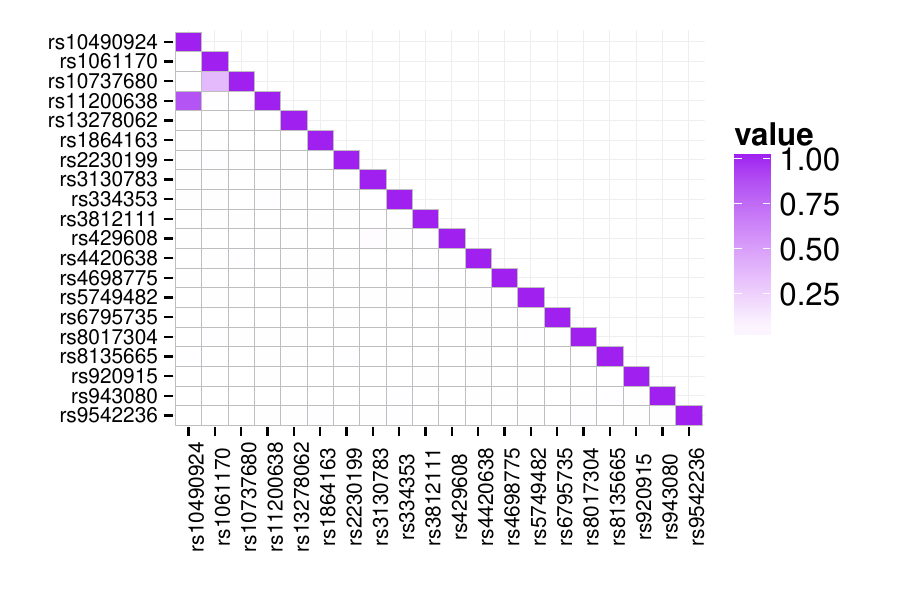}
\end{tabular}
\end{figure}

The contention that removal of   predictors that are essentially predictive of
each other is common in practice under the assumption of linkage disequilibrium
(LD) (i.e. nonzero association between SNPs). For example, SNPs
\texttt{rs10490924} and \texttt{rs11200638} in the
\textit{ARMS2/HTRA1} region are highly correlated (Figure
\ref{Fig:coinclusion}, right). However, our methodology suggests
that the inclusion of
either SNP does not automatically lead to the exclusion of the other. Actually,
although \texttt{rs11200638} has smaller  marginal II value compared to that of
\texttt{rs10490924}, Figure \ref{Fig:coinclusion} (left) shows that the two SNPs
are important in explaining AMD occurrence when they are observed
simultaneously, suggesting that there may be more unexplained factors at
play.

In the conditional inclusion importance graph, of note is the centrality of the ApoE gene.
Changes in this gene are known to be related to AMD. It is presently unknown what the role of ApoE is in relation to the
functional mechanism of effect with respect to allelic variation, although it is widely known that allelic variation can lead to significantly increased risk of
risk development of the aforementioned diseases. Figure \ref{Fig:coinclusion} 
suggest that multiple genetic variants together with ApoE are intertwined
with disease etiology.

\subsection{A comparison with previously reported and newly discovered
loci.} Columns 1--3 in Table \ref{table:consort} list newly discovered and
established SNPs. Columns 4--5 compare the SNPs newly discovered by the AMD Gene
Consortium
with the results of our ranking method using the 95\% confidence
level. It is reassuring that our approach identifies most of the well-replicated SNP
predictors, confirming the validity of our approach. Specifically, for the 12
well-replicated SNP variants, all except SNP \texttt{rs920915} show non-zero
inclusion importance, largely confirming the current consortium list. In addition,
our method finds two SNPs not currently part of the current consortium list (\texttt{rs1061170} and
\texttt{rs11200638} with respectable II values ranking 5th and 11th). These are
well known SNPs reputed
important by other studies corresponding to the CFH and
HTRA1 genes. Of the 6 newly found SNPs, only two receive non-zero inclusion
importance, meaning that there is little evidence in these data to suggest that
the remaining 4  SNPs contribute much information to explain AMD
occurrence beyond that already provided by the other SNPs with positive
inclusion
importance values.

\paragraph{Model combining and variable selection.} It may be reasoned that the SNP predictors with highest II values are
also the most pertinent to disease etiology. However,
there is no total presence of any single SNP in every LBM model. The
implication of a lack of unanimity points toward multiple possible disease pathways
and highlights the variable selection uncertainty faced in generating AMD risk
models. Nonetheless, a collection of stable SNPs can be obtained using
the aggregation strategy proposed in Section \ref{Sec:ModelAggregation}.

Table \ref{table:consort} (Columns 6--9) shows the result of our
LBM-aggregation
approach via AIC and BIC dimension reduction (AIC-$\LBM$ and BIC-$\LBM$).
For illustration purposes, we also report the selection obtained by other
popular model-selection methods. First, note that none of the SNPs with zero II
values are selected by the considered variable selection methods, confirming
the stability of our ranking scheme. Second, we illustrate that some of the
models selected by standard methods may miss some of the important SNPs at the
95\% confidence level. Specifically, the
LRT p-value for the BIC model selected by forward search (F-BIC)
is smaller than $\alpha=0.05$. This
means that the F-BIC model is outside the VSCS and should be regarded as overly
parsimonious since it is likely to miss one or more important SNPs related to
AMD.

 \begin{table}
\caption{SNPs selection and ranking for the AMD gene consortium data grouped in
previously reported (Rows 1--14)
and newly discovered (rows 15--20) SNPs in the AMD literature.
Columns 1--3: SNP, chromosome and gene identifiers. Column 4: SNPs previously
identified  as important. Column 5: Marginal
inclusion
importance statistic defined in (\ref{II}). Columns 6--7: Models selected by the LBM-aggregation strategy described in Section \ref{Sec:ModelAggregation} with
number of variables, $\widetilde{k}$, selected via AIC and BIC ( AIC-$\LBM$ and
BIC-$\LBM$). Columns: 8--9: Models selected by AIC and BIC step-wise forward
search. The
last row denotes the p-value from the LRT comparing the full model with 20 predictors with the selected models.\label{table:consort} }
 \rotatebox{90}{
 \begin{scriptsize} 
\begin{tabular}{lllccccccccc}\hline
{SNP id}                 & {Chromosome} & {Gene} & {Consortium}  &
\emph{II} & \textbf{AIC-$\LBM$}&\textbf{BIC-$\LBM$} & {F-AIC} & {F-BIC} & LASSO
& SCAD & MCP \\
\hline
\\
\multicolumn{12}{c}{Loci previously reported loci with marginal $\text{p-values} < 5 \times
10^{-8}$:}        \\ \\
\textit{rs10490924} & 10         & \textit{ARMS2/HTRA1}      & \checkmark
    & 0.91 & \checkmark & \checkmark & \checkmark & \checkmark &\checkmark &\checkmark&\checkmark \\
\textit{rs1061170}  & 1          & \textit{CFH}              &              &
0.66     & \checkmark  &   \checkmark         &\checkmark  &   &\checkmark  & &      \\
\textit{rs10737680} & 1          & \textit{CFH}              & \checkmark
&0.77    &\checkmark   & \checkmark & \checkmark & \checkmark &\checkmark&\checkmark&\checkmark\\
\textit{rs11200638} & 10         & \textit{ARMS2/HTRA1}      &              &0.35    &             &             &\checkmark  &   &  &       \\
\textit{rs13278062} & 8          & \textit{TNFRSF10A}        & \checkmark
    &0.19         & & &&&\checkmark&\checkmark&\checkmark\\
\textit{rs1864163}  & 16         & \textit{CETP}             & \checkmark
    & 0.18   & & & &&&&\\
\textit{rs2230199}  & 19         & \textit{C3}               & \checkmark
    & 0.59   & \checkmark &       &\checkmark        &     &\checkmark&\checkmark &\checkmark\\
\textit{rs3812111}  & 6          & \textit{COL10A1}          & \checkmark
    & 0.30     &&&&&\checkmark&\checkmark&\\
\textit{rs429608}   & 6          & \textit{C2/CFB}           & \checkmark
    & 0.06     &&&&&\checkmark&&\\
\textit{rs4420638}  & 19         & \textit{APOE}             & \checkmark
    & 0.73    &\checkmark & \checkmark &&&\checkmark&\checkmark&\checkmark\\
\textit{rs4698775}  & 4          & \textit{CFI}              & \checkmark
    & 0.48    &\checkmark  &&\checkmark&&\checkmark&\checkmark&\checkmark\\
\textit{rs5749482}  & 22         & \textit{TIMP3}            & \checkmark
    & 0.59     &\checkmark &&&&\checkmark&\checkmark&\checkmark\\
\textit{rs920915}   & 15         & \textit{LIPC}             & \checkmark
    & 0.00             & &&&&\checkmark&&\\
\textit{rs943080}   & 6          & \textit{VEGFA}            & \checkmark
    & 0.56       &\checkmark  &   & \checkmark &&\checkmark&\checkmark&\checkmark\\

    \\
\multicolumn{12}{c}{Newly discovered loci with marginal $\text{p-values} < 5 \times 10^{-8}$:}
                         \\ \\
\textit{rs3130783}  & 6          & \textit{IER3/DDR1}        & \checkmark
    & 0.73 &  \checkmark&    \checkmark       &\checkmark& &\checkmark&\checkmark&\checkmark\\
\textit{rs8135665}  & 22         & \textit{SLC16A8}          & \checkmark
    & 0.47  &&&& &\checkmark &\checkmark&\checkmark\\
\textit{rs334353}   & 9          & \textit{TGFBR1}           & \checkmark
    &0.00           &&&&&&&\\
\textit{rs8017304}  & 14         & \textit{RAD51B}           & \checkmark
    &0.00             &&&&&\checkmark&&\\
\textit{rs6795735}  & 3          & \textit{ADAMTS9/MIR548A2} & \checkmark
    &0.00             &&&&&\checkmark&&\\
\textit{rs9542236}  & 13         & \textit{B3GALTL}          & \checkmark
    &0.00           &&&& &\checkmark&&\\
    \\
\multicolumn{5}{l}{\emph{LRT p-value =}} &{0.6913} &{0.0625}&{0.5692}&{0.00037
}&0.7950&0.2832&0.2583\\
\hline
\end{tabular}
\end{scriptsize}}
 \end{table}

\section{Conclusion and final remarks}

Variable selection uncertainty is a very common but often overlooked issue in
applications of generalized linear models. In case-control genetic
studies, the common practice of explaining a phenotype in terms of a
single SNP combination may be inadequate when small signal-to-noise ratio 
prevents one
to learn whether the selected combination of predictors is the
true explanation behind disease. As a consequence, often multiple SNP
combinations are found to be compatible with a sample depending on the level of
noise, statistical model adopted, and variable-selection method. To address this
model-selection ambiguity, we extended previous work by \cite{ferrari15} and
introduced the notion of variable selection confidence set (VSCS) based on LRT
screening for GLMs. We applied the new approach to rank genetic factors
underpinning AMD -- an important eye disease with high prevalence
in the global population --  obtaining a stable
selection of SNPs based on available AMD case-control data.

The VSCS contains models  statistically
equivalent to the true model at a
pre-specified confidence level (e.g. 99\% or 95\%), thus representing a natural
extension of the familiar confidence intervals for parameter estimation in the
variable selection setting. We evaluated the validity of VSCSs constructed by
LRT screening using simulated SNP data
where the signal is not so clear, which we believe is very
common in complex diseases.  With adequate sample size, our results indicate
that the coverage probability of the true model is near the nominal
confidence level. Our numerical findings
confirm earlier results by  \cite{ferrari15} in the context of  linear
regression and suggest that the cardinality and composition of the lower
boundary models can be used to assess both variable selection
uncertainty and learn the true model structure.

The VSCS methodology provides the practitioner with new tools in support of the variable
selection activity in a way that goes beyond trusting a single selected model.
Here we  advocate the importance of a special subset of the VSCS, the lower
boundary model set, $\LBM$ (i.e. the set
of maximally parsimonious models surviving the LRT screening). Our numerical
examples confirm that, collectively, the
models in $\LBM$ contain a wealth of information on the true underlying model structure which can be used for variable ranking. Specifically, the
frequency of  predictors appearing in the LBMs (i.e. the II statistic defined in (\ref{II})) is a natural measure of their relevance
in explaining the response. For example, a predictor with II value near 1
plays a role in most of the parsimonious and
plausible explanations of the data. Ranking predictors based on the II
statistics is found to be a reliable indicator of whether a given regression
term is contained in the true model when the signal is small compared to noise. 
Similarly,
the joint (conditional) frequency of two or more predictors in the LMBs is
informative on their joint (conditional)
importance for explaining the response. In Sections \ref{Sec:importance} and
\ref{sec:real_data} we explained how this information can be effectively
conveyed
visually, for example using common directed graphs.

Another contribution of this paper is a new approach for selecting a single
central model based on the II ranks of predictors (Section
\ref{Sec:ModelAggregation}). Our Monte Carlo simulations suggest that model 
aggregation by
II-ranking is potentially a rather effective model-selection strategy in its 
own right
typically outperforming other established model-selection approaches that do not
benefit from selection-stability control. Due to these promising results, we
believe that developing a  theoretical understanding of the  properties
of this new aggregation approach in the future,  including optimal selection of
the reduced model size ($\widetilde{k}$), would be very valuable.

In the AMD Genotype data application, various models selected by
certain classic variable selection methods are suboptimal in finding SNPs
that are known to explain AMD etiology. Particularly, aggressive
selection methods promoting very sparse models such as BIC are found to fall
outside the VSCS. The usefulness of the proposed
statistical framework for novel variant investigation is shown in
the identification of SNPs in a set of well-replicated variants. The
lower boundary set of models was found to identify the vast majority of
well-replicated variants presented by the AMD Gene Consortium.
In novel variant investigation, the VSCS aids by screening out newly found SNPs by
genome-wide scans that do not contribute much information when considered
together with well-established SNPs.

In the AMD Genotype data  motivating this paper, the total number
of SNPs in the full model is relatively small ($p=20$). Thus, our methodology
is adequately handled by generation of the entire confidence set and
exhaustive LRT screening. Clearly, the exponential increase in models per
additional SNP forces alternate methods more efficient at deriving a
representative confidence and lower boundary sets. To address this issue, in our
future work, we plan to extend the current procedure and develop a screening
approach based on test statistic designed to handle a large number of
predictors. These may be complemented by computationally efficient methods that
can handle computation on larger model spaces.  For example, we
believe that certain Markov Chain Monte Carlo methods such as the reworked Gibbs
sampler described in \cite{qian2002using} may be used to overcome these
computational issues.

\bibliographystyle{rss}

\end{document}